\documentclass[11pt, a4paper]{scrartcl}
\usepackage[numbers,sort&compress]{natbib} 
\usepackage{graphicx} 

\begin{document}
\sloppy

\newcommand{\pref}[1]{figure~\ref{#1}}
\newcommand{\e}{\textrm{e}}
\renewcommand{\textsc}[1]{#1}
\newcommand{\includefigure}[2]{
  \begin{figure}[!htb]
    \begin{center}
      \includegraphics[width=\linewidth,angle=0]{#1}
      \caption{#2}
    \label{#1}
    \end{center}
  \end{figure}
}

\author{
  Felix P\"{u}tsch \textless{}fxp@thp.uni-koeln.de\textgreater\\
  Institute of Theoretical Physics\\
  University of Cologne\\
  D-50923 Cologne, Germany
}
\title{Analysis and modeling of science collaboration networks}

\maketitle
\begin{abstract}
  \small
  We analyze a science collaboration network, i.e.\ a network whose
  nodes are scientists with edges connecting them for each paper
  published together. Furthermore we develop a model
  for the simulation of discontiguous small-world networks
  that shows good coherence with the empirical data.
  
\end{abstract}


\section{Introduction}
  Hearing the term \emph{network}, the first association coming into
  one's mind are physically wired networks as telephone or computer
  nets.
  
  However, \emph{network} does also denominate the same concept on an
  universal level: nodes of whatever type connected by links determined
  by relations of the most different kinds. Mathematically spoken, we
  often call networks \emph{graphs}, nodes \emph{vertices} and links
  \emph{edges}.
  
  Thus, there are numerable different kinds of networks, physical
  ones (e.g.\ hard wired) as well as logical (e.g.\ dependencies)
  or social ones (e.g.\ contacts, friendships), stretching out to
  topics far from wired networks \cite{eb-mi02, li-ed01}.
  The area is under vigorous research.
  Good reviews can be found in
  \cite{al-ba02, do-me02, barabasi02, do-me03}.
  
  As a crucial difference from true random graphs \cite{so-ra51, er-re59},
  Lots of human created networks show the \emph{small world} effect
  \cite{wa-str98, watts99, newman00}, i.e.\ the average path length between two
  random points is significantly shorter (behaves logarithmically with
  the system size) than that of random networks (linear with system size).
  
  A model developed by \textsc{Barab\'{a}si} and \textsc{Albert} \cite{ba-al99a}
  describes several small-world networks very well, included e.g.\ the
  world wide web \cite{al-je99, ba-al00}.

  In the context of science the network between
  scientists as nodes of the graph is of particular interest.
  This network belongs to the group of social ones, with
  humans as nodes.
  Unlike most other forms of social
  relationships, that are quite difficult to
  capture objectively, the field of published papers is very
  widespread covered by the \emph{Science Citation Index} \cite{SSI}
  and so easily available to research.
  

\section{Collaboration networks}
\subsection{Typology}
  Apart from linking not the scientists but their papers
  \cite{raan90},
  there are basically two possible choices in what to consider a link
  between two authors---both covered equally by the database.
  \begin{enumerate}
  \item We chose to consider citations from one author to another as
    links \cite{redner98}.
    In this case we get a rapidly growing number of papers we have
    to consider, as each added scientist cites several others and so on.
    It even is not clear \emph{a priori} that we ever get to an
    end---except in the case of having \emph{all} scientists in our
    set of authors.
  \item The only connections in our net are those of co-authorship in
    one or several papers \cite{newman01a,newman01b}.
    Now we have the advantage of
    being able to chose an arbitrary set of authors as start of our
    examinations---though we should provide a preferably
    reasonable one. This choice will be discussed later.

    But another problem
    arises using this method: it is very difficult---if not
    impossible---to determine \emph{all} papers a given pair of
    authors ever published.
  \end{enumerate}

\subsection{Building the net}
  \label{building}
  As a solution we chose the following proceeding: We start with one paper. As
  one part of our work will deal with \textsc{Barab\'{a}si-Albert}
  networks, we take the corresponding paper \cite{ba-al99a} as center of
  our investigation.

  To determine the set of authors we want to deal with we select all
  185 papers that cite this paper. (Remark: we have to be careful not to
  mix citation data from different dates as new papers are continuously
  added to the database. The base of our investigation is October
  21\textsuperscript{st}, 2002.)

  As a second step we construct a list of unique authors from all these
  papers. In a first approach we have 559 scientists, whereof some
  turn out to be identical only appearing in particular papers with typos.
  We finish with a set of 555 authors to whom we attribute consecutive numbers.

  The last step of the network creation consists in establishing links
  between all these authors. This is done by selecting one paper after
  the other and introducing a connection between each possible pair of
  this paper's authors.

  Eventually, this gives us a net which we suppose to be rather typical for
  scientific collaboration.

  The network size is relatively small compared to all data in the
  \emph{Science Citation Index} (approx. $10^7$ papers).
  We will study properties
  of this subnet and compare it to some classical and recent network
  models hoping to get an idea of what leads to the structure we
  observe. Verification with bigger networks is a task
  for the future.


\subsection{Statistics}
  Now we will analyze a crucial property of the just created net: the
  cluster size distribution. In our case, clusters of scientists are formed 
  by the links between them, i.e.\ a single paper of $n$ authors
  already forms a cluster of size $n$.
  
  Immediately, our eyes are caught by a
  paper on the \emph{Human Genome Project} \cite{venter01} with 274\
  authors. The giant cluster thus formed is singled out from all others
  by its hugeness. Regarding it as an anomaly, we chose to remove it
  from our network.
  A brief examination yields that this is no harm as the scientists
  participating in this work did not cooperate with the others
  of our study and form a
  big cluster containing only themselves.
  We will discuss later in section \ref{explanation_giant}
  if this proceeding was justified.
  
  Now, we investigate the frequency of clusters
  of a given size. Our expectation is to see many clusters with few
  authors and \textsl{vice versa}.

  \includefigure{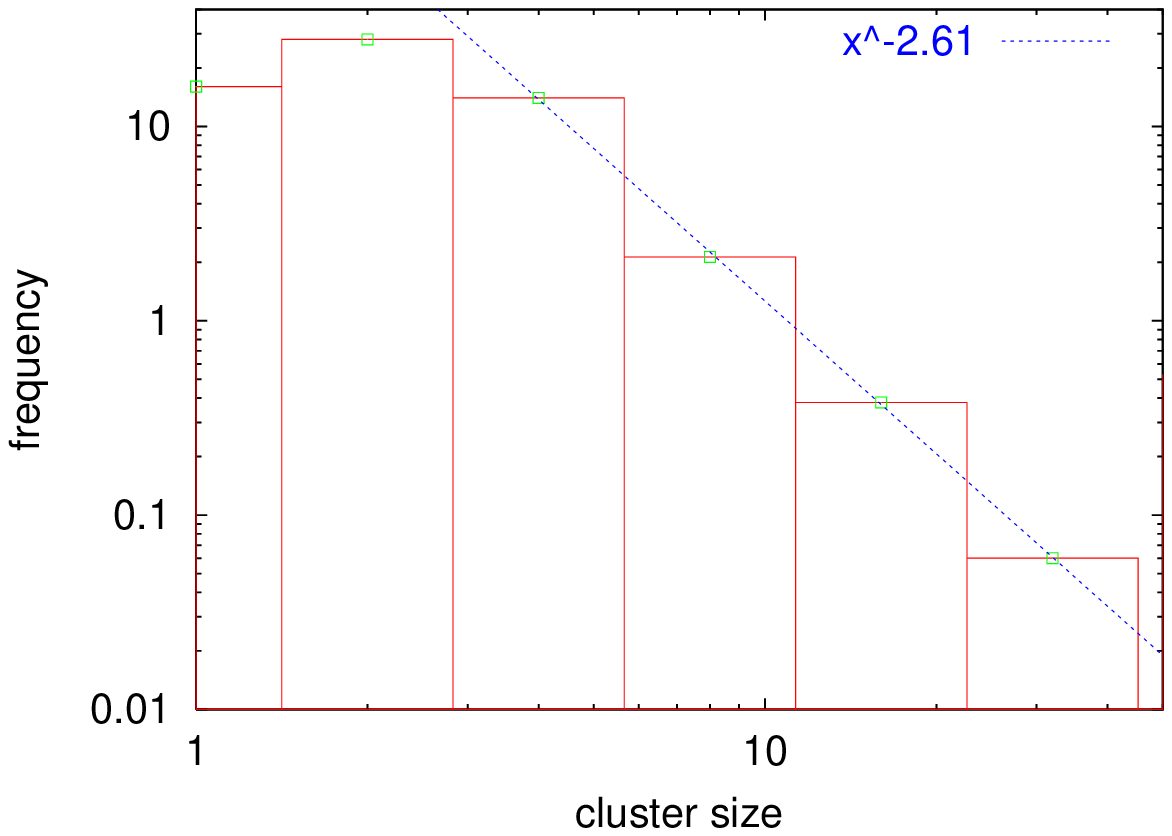}
    {frequency distribution of the cluster size}

  The experimental data (\pref{puetsch2}) shows this behavior but with one
  surprise: although the most frequent cluster size is $2$ due to
  a big number of publications with two authors, most
  scientists maintain collaboration with three others.

  A probable explanation is this: scientists are often member of
  research groups involved in different themes, thus connecting different
  clusters formed by single two-author-papers.


\section{Computer simulation}
  We now describe a model reproducing these results,
  hoping to understand how this macroscopic
  behaviour arises by microsopic decisions of the individuals.
  
  A model describing small-world networks quite well is that
  of \citet{ba-al99a}. Unfortunately, it only deals with networks
  consisting of one single component. We generalize it
  to discontiguos networks.
  
\subsection{Standard \textsc{Barab\'{a}si-Albert} model}
  The \textsc{Barab\'{a}si-Albert} network model
  starts with a set of (mostly) $m_0=3$ points, each connected to each
  other by a link.
  
  In each time step, (usually) $m=3$ new nodes are added, each with a
  link to the existing network. The probability of an already existing
  vertex to be a linking target is proportional to the number of
  connections already present at this node. ``The rich get richer.''
  
\subsection{Modified model}
  To cope with networks consisting of several components, we have to
  modify the model. We chose a very simple approach: In each step of
  adding nodes we start a new network of $m_0=3$ nodes with at certain
  probability $p$.
  
  Vertices added in consecutive time steps can connect to any node in
  any component respecting the same probability rule as in the standard
  model.
  
  In the case $m=1$, components can only grow (\emph{isolated clusters}),
  whereas
  in the case $m>1$, new nodes are able to connect two or more existing
  components of the network (\emph{merging clusters}).
  
\subsection{Results}

  To compare the results with the data collected in section \ref{building},
  we let the network grow up to the
  same size of $555$ nodes. This is repeated $10^4$ times for statistical
  reasons.

\subsubsection{Isolated clusters}

  In the case $m=1$, i.e.\ the case \emph{isolated clusters}, we can be sure
  to get scale-free behavior within the distinct clusters, as
  the probabilities for attachment of a new node to an existing one are
  the same as in a single \textsc{Barab\'{a}si-Albert} network (modulo
  a proportionality factor due to a new node having a ``choice'' between
  different clusters to connect to).
  
  However, the complete network does not necessarily have to be scale-free,
  as the total statistics is a sum of multiple scale-free sub-networks
  or clusters.

  \includefigure{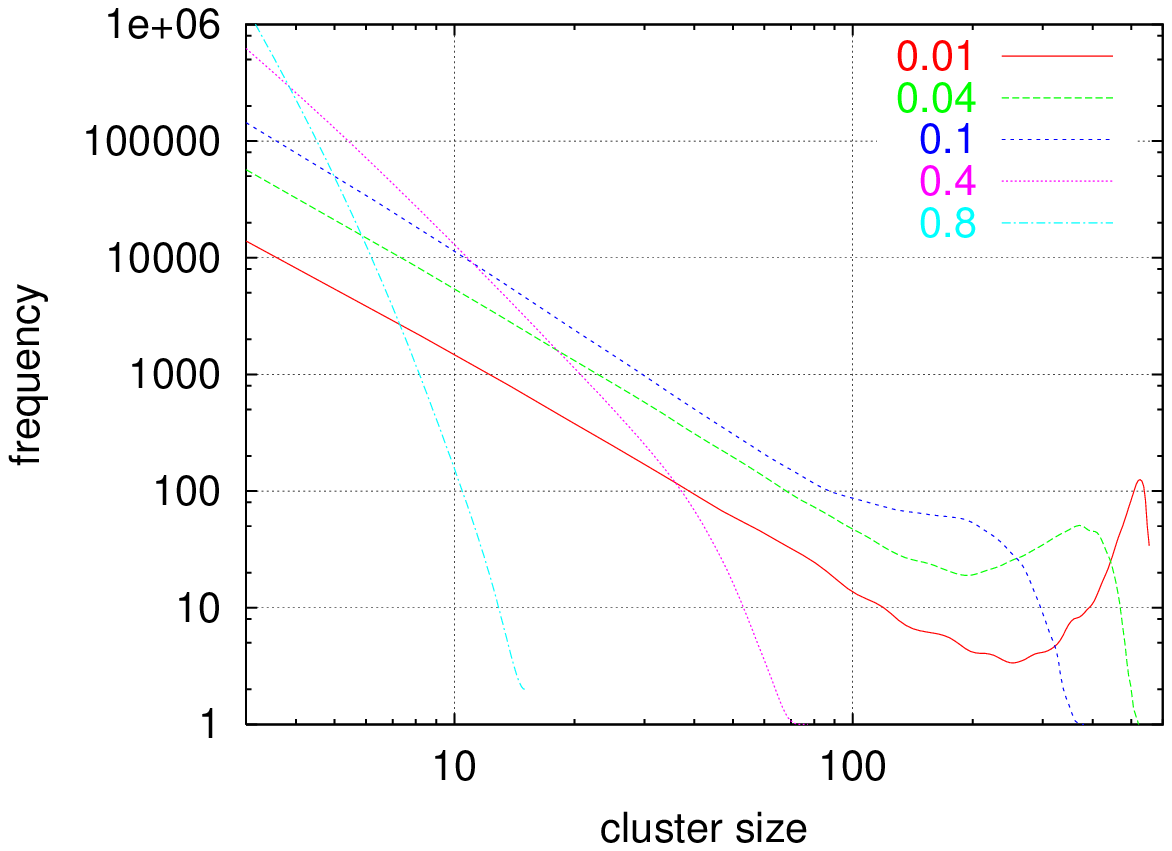}
    {Frequency of clusters vs.\ cluster size at different
      probabilities for a new net. Simulation was run $10^{4}$ times
      with a network growing up to $555$ nodes. The curve for $p=0.01$
      is the one with the rightmost peak; to the left follow the other
      $p$-values in ascending order.}
  
  First, we examine the number of clusters of different sizes
  (\pref{puetsch3}).
  
  Obviously, a high probability of starting a
  new net leads to many smaller networks, whereas a low one privileges
  bigger networks. Yet, we make an interesting observation: low
  probabilities lead to a cluster-size distribution that is not
  monotonic any more, but favors big networks.

  \label{explanation_giant}
  The explanation is straight-forward:
  For $p=0$ we will see a graph $\propto\delta(555)$, as there
  is only one giant cluster, for
  $p=1$ a graph $\propto\delta(m_0=3)$, because there are only embryonic
  sub-nets. What we observe for $0<p<1$ is
  the transition between both extremes.
  
  For all $p$ we start with a power law region regarding the distribution
  for small and medium cluster sizes. The exponent varies with the
  network-birth probability $p$. In \pref{puetsch4} we analyze this
  correlation in a semi-logarithmic plot.

  \includefigure{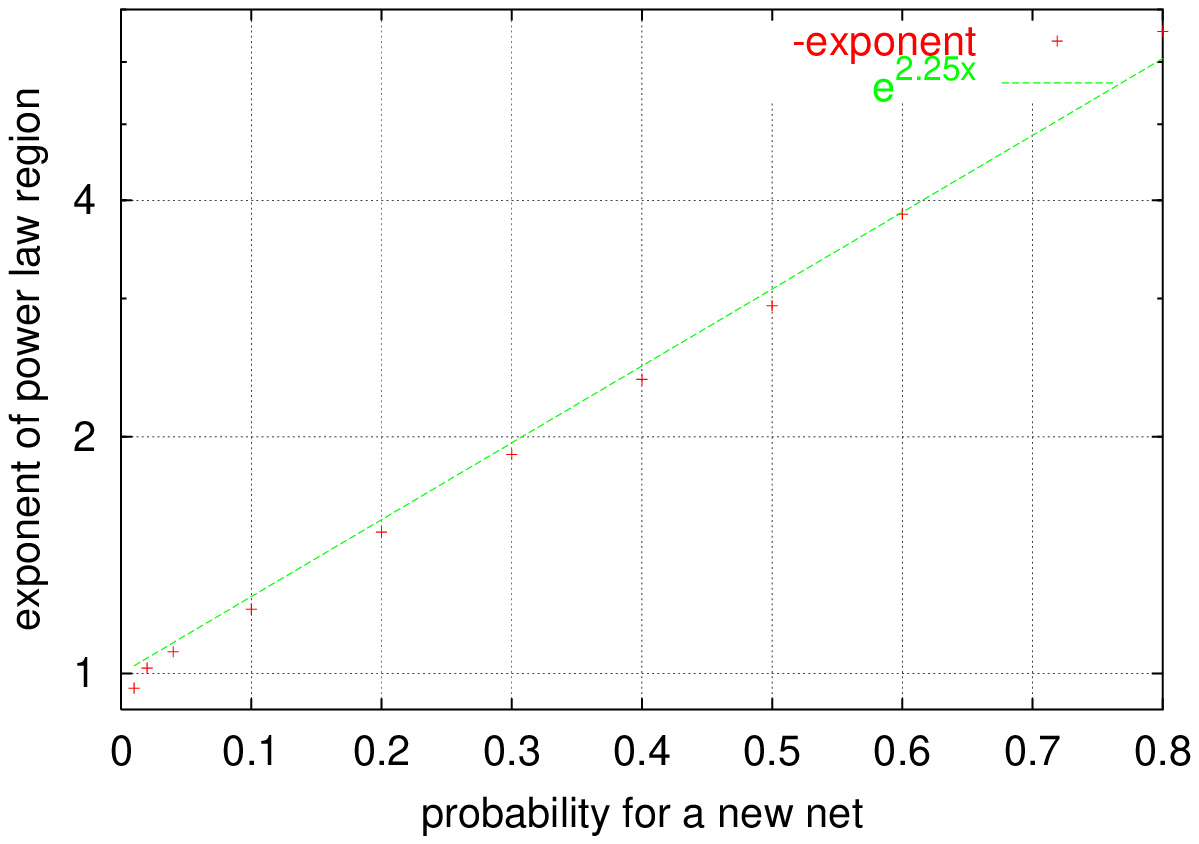}
    {Negative exponent of the power law
      part of the curves in \pref{puetsch3} vs.\ probability $p$
      for a new net. The line corresponds to $exponent=-\e^{2.25p}$.}
  
  We find that $-\e^{2.25p}$ describes our data rather well.
  Of course, this formula cannot be true for general $p$ as for
  $p\to 1$ we expect $m\to -\infty$!
  
  Finally, we compare the cluster distribution from computer simulation
  with real-world data (\pref{puetsch5}) and are surprised.
  The data fits well---including the giant cluster we thought to
  be an anomaly whilst building the network. The plot gives strong
  evidence that indeed it was an organic part of the network. Its
  hugeness is simply due to the graph forming rules.
  
  \includefigure{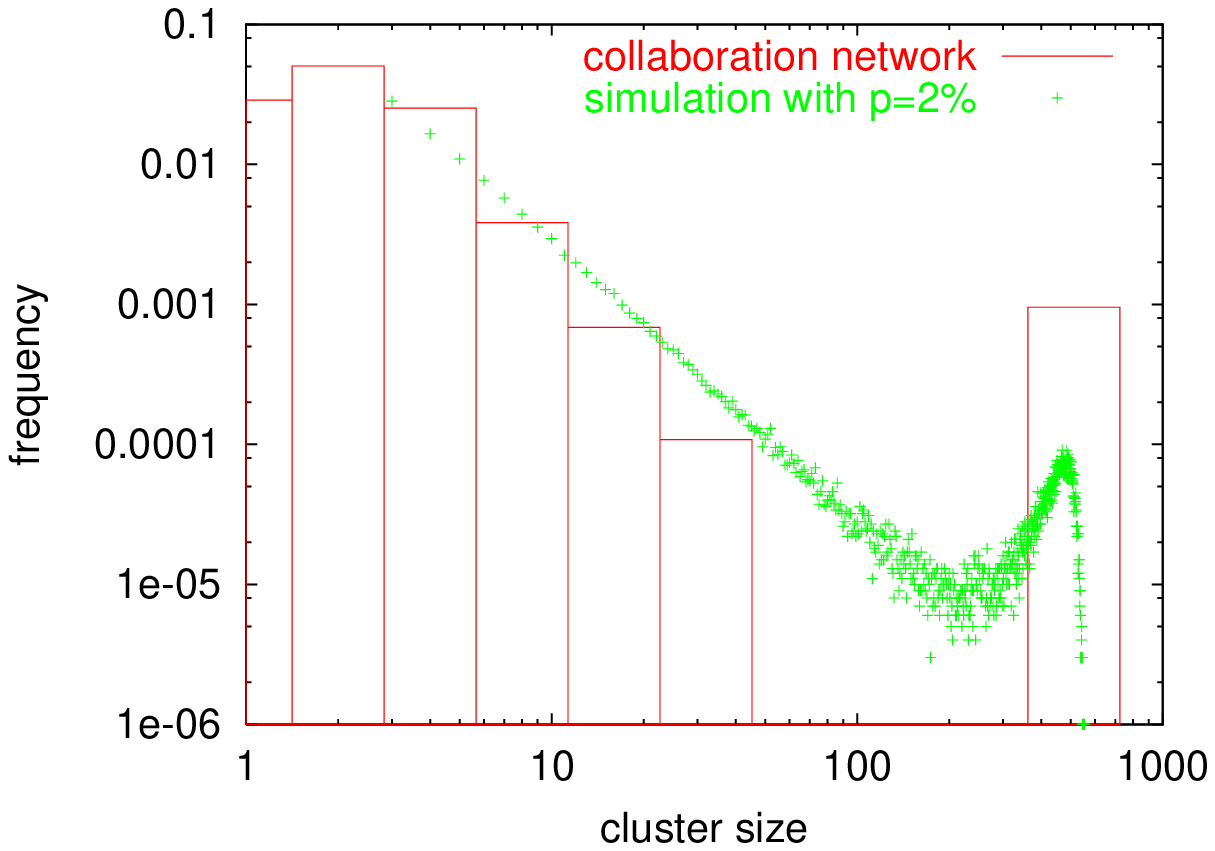}
    {Comparison of the simulation with $p=0.02$ using the \emph{isolated
      clusters} model of \pref{puetsch3} and statistical data
      from a science collaboration network. To simplify
      comparison, relative frequencies are used.}

\subsubsection{Merging clusters}

  Now, we modify the model by examining $m>1$. In this case, newly
  added vertices develop several links to existing nodes (and thus
  existing clusters), being able to connect hitherto separated
  networks. In this paper, we limit our considerations to the
  standard \textsc{Barab\'{a}si-Albert} case $m=m_0=3$.
  
  Using different $p$, we quickly recognize
  that low and medium probabilities make the simulation
  nearly always end up with
  a single giant cluster containing all vertices.
  Points of interest are higher $p$ in the region of $60$--$90\%$.
  
  \includefigure{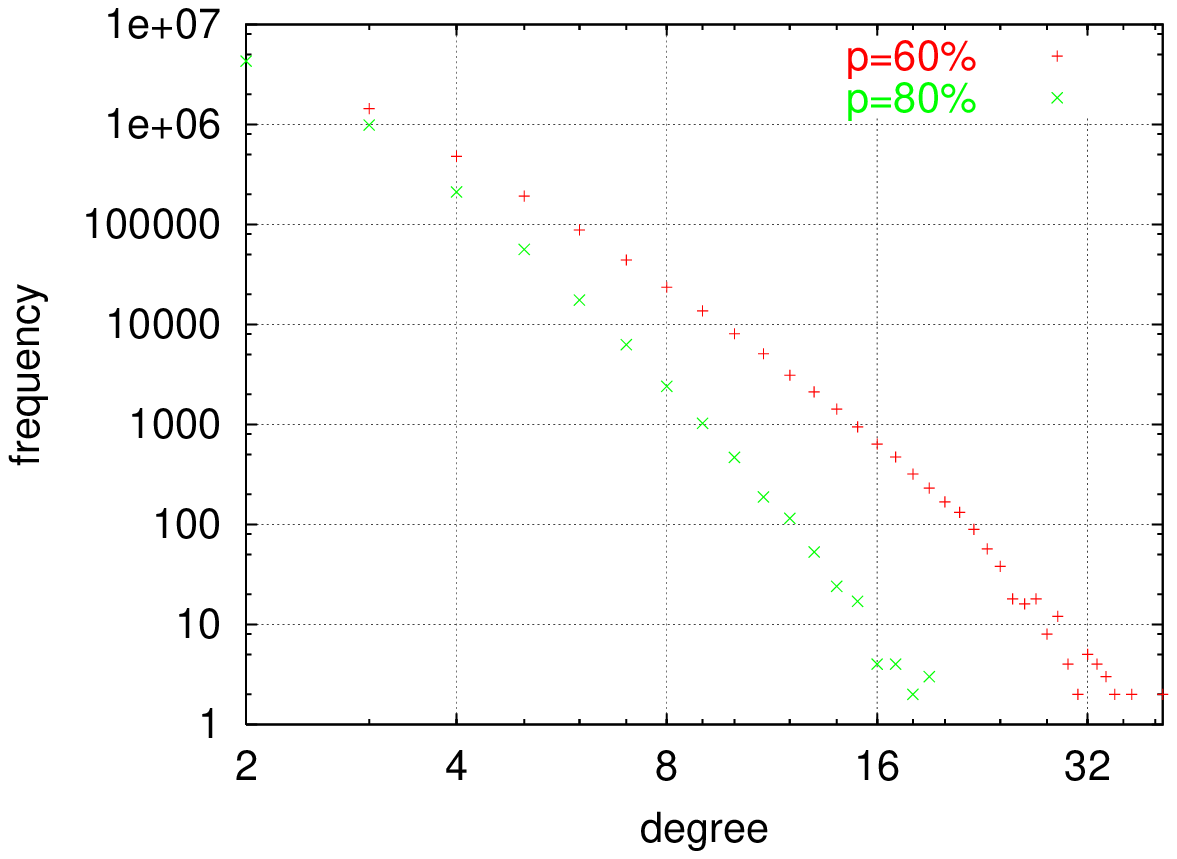}
    {Frequency of nodes with a certain degree.
      Simulation was run $10^{4}$ times
      with a network growing up to $555$ nodes.}

  What about scale-free behavior in this case? In \pref{puetsch6}
  we can see that there is no pure scale-free behavior. There seems
  to be power-law behavior for small degrees and an exponential
  cutoff at higher values as has been verified by a semi-logarithmic
  plot of \pref{puetsch6}.
  Similar results have been observed by
  \citet{newman01a} for collaboration networks.
  
  One could argue that this effect is due to the fact that we
  do not plot the degree distribution for single clusters but for the
  whole set of them. This demur only counts at first sight, though.
  At $p=80\%$ we have several small clusters but virtually only one
  giant cluster dominating the degree distribution for high degrees.
  So, the fact of averaging of many different sized clusters should
  manifest mainly in the area of small degrees opposite to our observations.

  For small cluster sizes, we observe a non-uniform behavior
  regarding the frequency of clusters of a given size. There is no
  monotony of the sort that larger clusters are less probable than
  small ones.
  
  The explanation is as follows: newly born clusters have a size of
  $m_0=3$ and thus appear very often. Also, cluster of sizes $4$ or $7$ are
  very probable, whereas a cluster of size $5$ is very rare, because
  it can only be formed by a new cluster to which two new ones have
  connected without glueing it to a second cluster.
  
  In a semi-logarithmic plot, we find a parabolic dependence for high
  cluster sizes (i.\ e. a \textsc{Gauss}ian distribution around a mean
  depending on $p$).
  
\section{Conclusion}

  We constructed a network of coauthership with $555$ authors. Only scientists
  were chosen that cite a specific paper \cite{ba-al99a}. We find a
  cluster size distribution showing an exponential decay for small cluster
  sizes and a giant cluster that cannot be explained by common network
  models.
	
  A change of the model of \citet{ba-al99a} by
  allowing a certain probability to start new clusters, enables us to
  simulate networks consisting of distinct clusters, e.g.\ friendship
  or collaborational networks.
  
  The modification provides two different models. In the \emph{isolated
  clusters} variant, newly born clusters stay distinct forever and
  show up scale-free behavior on their own.
  
  This model is able to explain facts formerly
  regarded as statistical anomalies as the observation of a giant
  cluster of a size exceeding largely all others in the network.
  Following the simulation this observation fits very well
  (\pref{puetsch5}).

  \emph{Merging clusters} is the second variant, i.e.\ new nodes are able
  to merge existing clusters. This model shows an exponential fall-off
  for higher degrees in the degree distribution and thus no pure
  scale-free behavior. 
  Furthermore, it results mainly in a
  \textsc{Gauss}sian distribution of cluster sizes for bigger clusters
  and thus cannot cope with reality.
  
  More results will be given in \cite{puetsch03b}.

\section*{Acknowledgements}

I would like to thank D.~Stauffer
for several ideas, comments and discussions on this paper.


\end{document}